\documentclass{article}    

\usepackage{graphicx}
\begin{document}

\title{Dynamic modeling of gene expression in prokaryotes: application to glucose-lactose diauxie in \textit{Escherichia coli}}
\author{Jaroslav Albert, Marianne Rooman\\ \\BioSystems, BioModeling and BioProcesses,\\ Universit\'e Libre de
 Bruxelles, CP165/61,\\av. Roosevelt 50, 1050 Bruxelles, Belgium}
\date{}
\maketitle

\begin{abstract}
Coexpression of genes or, more generally, similarity in the expression profiles
poses an unsurmountable obstacle to inferring the gene regulatory network
(GRN) based solely on data from DNA microarray time series. Clustering of genes with similar expression profiles
allows for a  course-grained
view of the GRN and a probabilistic determination
of the connectivity among the clusters. We present a model for the
temporal evolution of a gene cluster network which takes into account interactions of gene products with
genes and, through a non-constant degradation rate, with other gene products. The number of model parameters is reduced  by using polynomial functions
to interpolate temporal data points. In this manner, the task of parameter estimation
is reduced to a system of linear algebraic equations, thus making the
computation time shorter by orders of magnitude. To eliminate irrelevant networks, we test each GRN for
stability with respect to parameter variations, and impose restrictions on its behavior near the steady state. 
We apply our model and methods
to DNA microarray time series' data collected on \textit{Escherichia coli} during glucose-lactose diauxie
and infer the most probable cluster network for different phases of the experiment.

{\em Keywords: Gene regulatory networks, DNA microarray, gene product degradation, gene clustering,
optimization, dynamic robustness}

\end{abstract}

\section{Introduction}
\label{intro}

The information encoded in the genome of
living organisms has presented a new level of complexity that continues to challenge
the mind. Both theoretical and experimental studies have proven to be very difficult
mainly due to the large dimension of the system of interacting
genes. Even the simplest of prokaryotic cells contain some 4000
genes of which a significant fraction participates directly or indirectly in regulating
(enhancing or inhibiting) the expression of one another.
Despite these daunting obstacles, progress in unraveling
Gene Regulatory Networks (GRN) has been made mainly due to
new experimental methods that allow for more detailed studies
of the intricate mechanisms within the living cells.

In particular, owing to the development of  DNA microarrays techniques, it has become
possible to probe the behavior of thousands of
genes simultaneously over a certain course of time.
The advantage of recording the temporal evolution of the genome (or at least a large part of it)
as compared to having the same information
on only a handful of genes is obviously quite important and has led to the
onset of new types of studies, \textit{e. g.} see
\cite{Gardner} and \cite{Bolouri}. The challenge
of inferring the GRN however still persists due to the abovementioned problem of
large dimensionality, but
also because of (still) partial and noisy data.
This makes improvements in both theory and
experiment equally important.

The difficulty of dealing with a large number of genes has forced one to proceed to various simplifications of the GRN problem.
In many studies researchers have made use of simple models for the gene-gene product (GP) interactions
including Boolean and Baysian networks \cite{Lahdesmaki}, and linear coupled differential equations
as in the works by \cite{Haye}, \cite{Gebert} and \cite{Liebermeister}.
Although simple models are attractive, especially when dealing with a large number of genes,
they often suffer from lack of physical relevance. The biochemical interactions on
the molecular level are known to be more complicated than most simple
models can account for. Nevertheless, the correlations between genes that many
simple models predict can provide   a course-grained view of the GRN.
Another popular trend in simplifying the GRN problem
is to focus on a subset of genes that are known to
regulate each other under the assumption
that no other gene
has a regulatory influence on this sub-network.
This type of approximation makes more
complicated models, \emph{i.e.} non-linear models,
feasible as was shown in \cite{Vohradsky} and \cite{Vu}.

An additional difficulty is related to the fact that many genes are coexpressed
and exhibit thus basically the same expression profiles, and that even non-coexpressed genes
may have similar profiles under certain circumstances and/or during a certain time span. As a consequence
one cannot obtain the information
about the correlations among all genes from
DNA microarray data alone, regardless
of the quality or complexity of the model at hand.
This issue can be partly resolved by grouping genes with similar expression profiles
into clusters, which allows one to shift focus from
individual genes to the study of how clusters
influence each other.
Consequently, the
dimension of the problem is reduced to the number of clusters,
which in many cases is much less than
a hundred.
Such a drastic reduction in dimensionality also
opens the door to more complicated (and hence more accurate)
models, and can reveal a more realistic course-grained picture
of the GRN. The last difficulty we address here is the
multitude of possible GRNs that generate the same
gene (or cluster) expression profile.
This is sometimes referred to as gene elasticity
\cite{Krishnan}.

In this paper we attempt to further the methods of identifying the connections
among gene clusters on the basis of DNA microarray time series and propose
criteria that eliminate many possible solutions for the gene cluster network.
The system under consideration consists of
\emph{E. coli} bacteria in a glucose-lactose environment. The GRN model we design
takes into account gene-GP and GP-GP interactions
that are derived from physical arguments. Our work is a further 
step towards reliably predicting cluster gene 
networks on the basis of DNA microarray time series.

The paper is structured as follows.
In section one we give a brief description of the known regulatory mechanisms in
prokaryotic cells and derive a physical model that describes them.
We then adapt this model to be suited for cluster-cluster interactions.
In section two we discuss the procedure of parameter identification and parameter
reduction. The application of our model to \emph{E. coli} during glucose-lactose diauxie is covered in section three where
we also discuss the criteria for GRN selection. In the last section, we summarize
our results and discuss possible issues as well as outlooks
for further studies in this field.

\section{Modeling the biochemical processes}
\label{sec:1}
The living cell is like a small factory whose
products (GPs), \emph{i.e.} RNA and proteins,
sustain it, allow it to divide, and even terminate its life.
The cell absorbs various chemicals from the environment and
uses them for many purposes. Some of them serve as
fuel to drive its internal machinery, others
are used for intra-cell or cell-to-cell signal propagation and many
other essential functions.
Genes and their promoter sequences
act as pieces of software that hold
the instructions for synthesizing
RNA molecules, some of which, \emph{i.e.} the (messenger) mRNAs, are then translated into
proteins; all RNAs and proteins are collectively referred to as GPs
and we will make no distinction between them in what follows.

The GPs that bind to regulatory promoter sites and hence regulate the transcription of
genes (direct regulation) are called transcription factors (TF).
Other GPs which are not TFs themselves but bind to TFs can also influence gene regulation
(indirect regulation).
In fact, virtually all GPs can play a role, either directly or indirectly, in the regulation of gene expression.
Depending on the external environment
some genes may be highly active (expressed) while others can have low output or
even be completely off. Which gene is expressed and when depends on
the abundance of specific GPs and their affinity
to bind the gene's regulatory sites or other GPs.
Since gene regulation is much more complicated in eukaryotes than
in prokaryotes (see \emph{e.g.} \cite{Lodish}) we concentrate
on the latter. Hence, the rest of this article deals
exclusively with prokaryotes.

\begin{figure}
\begin{center}
\begin{picture}(0,0)(0,0)
\end{picture}
\includegraphics[scale=.45]{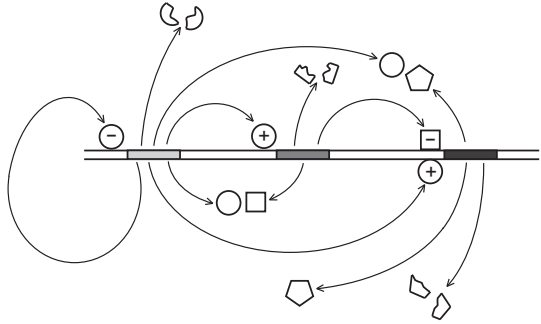}
\caption{A GRN of three genes showing the regulatory pathways. The circle, square, and
pentagon represent the GPs
synthesized by genes 1, 2, and 3, respectively (from left to right),
and the broken circle, square, and
pentagon represent degraded GPs.
The $+$ and $-$ signs correspond to activation and
repression respectively. The third gene
synthesizes a GP that is not a TF as it does
not directly regulate any of the other two genes. Also shown is
the interaction among the different GPs.}
\label{ThreeGenes}
\end{center}
\end{figure}

\subsection{A model of prokaryotic gene regulation}
\label{sec:2}
In figure 1. we show a toy model of three genes that mutually interact through
their GPs. The arrows
indicate the gene-GP and GP-GP interaction pathways. Because of
thermodynamic instabilities, degradation by enzymes, transport to other cell compartments,
and effects of dilution upon cell growth, GPs
inevitably degrade or loose activity in some characteristic time. Depending on this time GPs can have
either long or short lasting influence on genes. In what follows we present a
mathematical description
of how the GPs influence a gene's transcription rate.

We begin by assuming that time delays between the production of GPs and their influence on
a gene are negligible compared to the times over which the
concentration levels change significantly. This assumption is supported by comparing
the typical diffusion and transcription rates, and amounts to neglecting the
effects of translational regulation and any distinction between RNA and proteins.

The most general set of differential equations describing a
system of $M$ genes under constant environment
has the form
\begin{equation}\label{GeneralEq}
\dot{X}_i=R_i({\bf X},{\bf C}_i)
\end{equation}
where $i=1,...,M$. $R_i({\bf X},{\bf C}_i)$, ${\bf X}=(X_1,...,X_M)$ and ${\bf C}_i=(C^1_{i},...,C^N_i)$ 
are an influence function, the concentration  vector of GPs and the set of all parameters pertaining to gene $i$, 
respectively.
The influence function consists of two terms:
\begin{equation}\label{Rfg}
R_i({\bf X},{\bf C}_i)=f_i({\bf X},{\bf C}_i)+g_i({\bf X},{\bf C}_i),
\end{equation}
with
\begin{equation}
f_i({\bf X},{\bf C}_i)>0,\,\,\,\,\,\,\,\,g_i({\bf X},{\bf C}_i)<0.
\end{equation}
The first term in Eq. (\ref{Rfg}) depends on the abilities of
various GPs to bind a promoter of gene $i$, \textit{i. e.} the binding affinities, and on the
efficiency of recruiting the RNA-polymerase,
while the second term gives the rate of GP degradation. To understand the structure of
$f_i$, let us first look at a simple example of two TF's competing for the same promoter.
Suppose that the transcription of gene $i$ is enhanced by two activators $1$ and $2$.
The probability that say $1$ will bind to the promoter is (see \cite{Buchler})
\begin{equation}\label{Prob1}
P_A^1({\bf X},{\bf A}_i)=\frac{A_{i1} X_1}{1+A_{i1}X_1+A_{i2}X_2},
\end{equation}
with $X_1$ and $X_2$ being the concentrations of GP $1$ and $2$. The index $A$
stands for activation.
The parameters $A_{i1}$ and $A_{i2}$ are proportional to the
frequencies of
collisions between the gene's promoter and the GPs $1$ and $2$,
and to their binding affinities.
The term 1 in the denominator indicates that the promoter may be unoccupied, 
and the term $A_{i2}X_2$ comes from the fact that
the GPs $1$ and $2$ compete for the same promoter. Indeed, if the concentration of
GP $2$ becomes very large, the probability for GP $1$ to bind the promoter
becomes low. On the other hand, if the converse were to occur this probability
would approach one.

For a transcription to occur the part of the promoter 
which admits only activators must be
occupied while the part that admits repressors must be unoccupied. 
The probability for such a scenario
is given by the probability that an activator is bound to the
promoter multiplied by the probability that the converse is true for a repressor.
The probability for gene $i$ to have a certain rate of transcription
is then 
\begin{equation}\label{TR}
P_i({\bf X},{\bf\Gamma}_i)=P_A({\bf X},{\bf A}_i)(1-P_S({\bf X},{\bf S}_i)),
\end{equation}
where ${\bf \Gamma}_i=({\bf A}_i, {\bf S}_i)$, $A^j_i\geq0$, $S^j_i\geq0$,
\begin{equation}
P_A({\bf X},{\bf A}_i)
=\frac{A^j_i X_j}{1+A^j_i X_j}
\end{equation}
and
\begin{equation}
P_S({\bf X},{\bf S}_i)
=\frac{S^j_i X_j}{1+S^j_i X_j}.
\end{equation}
The repeated index $j$ sums over all the GPs which influence gene $i$: 
$A^j_iX_j=\sum_jA_{ij}X_j$. The total probability for gene $i$ to be occupied by any activator is then
\begin{equation}
P_i({\bf X},{\bf \Gamma}_i)
=\frac{A^j_i X_j}{(1+A^j_i X_j)(1+S^j_i X_j)}.
\end{equation}
Note that in deriving this equation we made the assumption that the expression
of a gene may be activated or repressed by a single GP, 
and does not require complexes of GPs or cascades of interacting GPs.
The motivation for this choice is
that most genes in prokaryotes are regulated by forming
DNA-protein complexes involving single proteins.

Since ${\bf X}$ is a stochastic variable we have to average
the transcription rate over an
ensemble of many cells under the same external conditions:
\begin{equation}\label{AvTR}
f_i({\bf X},{\bf\Gamma}_i)=\langle r_iP_i({\bf X},{\bf\Gamma}_i)\rangle.
\end{equation}
The parameter $r_i$ is the maximum transcription rate
corresponding to the saturation point at $A^j_i X_j\rightarrow\infty$
and is taken to be independent of
the particular combination of GPs that bind to gene $i$. Note that by averaging over an ensemble,
the stochastic variables of the system become determininstic. For that purpose we first write  in Eq. (\ref{AvTR}) $X_i=x_i+\eta_i$ where $x_i$ is the average expression of gene $i$ over an ensemble of identical cells and
$\eta_i$ is a Gaussian noise function of the same gene. We then
expand with respect to $\eta$:
\begin{equation}\label{fi}
f_i({\bf X},{\bf\Gamma}_i)=r_i\left[P_i({\bf x},{\bf\Gamma}_i)
+\frac{\partial}{\partial x_j}P_i({\bf x},{\bf\Gamma}_i)
\langle\eta_j\rangle+O(\langle\eta^2\rangle)+...\right].
\end{equation}
Since $\langle\eta_j\rangle=0$ for
a Gaussian function, the first order approximation of the rate function $f_i$ in Eq. (\ref{fi})
yields 
\begin{equation}\label{fi2}
f_i({\bf X},{\bf\Gamma}_i)=r_iP_i({\bf x},{\bf\Gamma}_i)=r_i\frac{A^j_i x_j}{(1+A^j_i x_j)(1+S^j_i x_j)}.
\end{equation}

Various environmental conditions within the cell can
cause the GPs to degrade, or loose their activity, after some characteristic time $\tau_c$
\cite{Maurizi}.
Since the ability of a GP to influence a gene
depends on how long it remains active, those GPs with
a long $\tau_c$ are more likely to bind a gene's promoter.
The converse is true for GPs with short $\tau_c$.
We hypothesize that GPs mutually interact to either prolong (\emph{e.g.}
through stabilizing complexes) or shorten (\emph{e.g.} through  degradation by proteases) their
$\tau_c$
in order to provide another channel for gene control.
The nature of GP-GP interaction is too complicated for our purposes
here and
will not be treated on the molecular level. Instead we want to write down
a course-grained expression that corresponds to the behavior expected from
the arguments just outlined.
In particular, we expect that overabundance of any one GP
would saturate its influence on other GPs. On the basis of this assumption,
we define
a general $\bf X$-dependent degradation rate $k_i$ of the form:
\begin{equation}\label{k}
\frac{1}{\tau_i}\equiv k_i({\bf X},{\bf \ell}_i)=\frac{K_{i}^{+}+K_{i}^{-}e^{K_i^jX_j}}{1+e^{K_i^jX_j}},
\end{equation}
where ${\bf \ell}_i=(K_i^{+},K_i^{-},{\bf K}_i)$, $K^{\pm}_i\geq0$, $-\infty<K^j_i<\infty$,
and $\tau_i$ is the characteristic time associated with GP i.
The two parameters $K_i^{+}$ and $K_i^{-}$ symbolize the maximum and minimum
degradation rate respectively. The matrix elements $K_i^j$ give the
influence of GP $j$ on GP $i$. Notice that when $K_i^jX_j$ is large and positive,
$k_i({\bf X},{\bf \ell}_i)$ approaches $K_i^{-}$ which corresponds to the longest $\tau_i$
for GP $i$ while the opposite limit yields $K_i^{+}$ - the shortest $\tau_i$.
The sign of each matrix element $K_i^j$ determines whether GP $j$ has a stabilizing
(a plus sign) or destabilizing (a minus sign) influence on GP $i$. As
before, Eq. (\ref{k}) must be averaged over the ensemble of cells. 
The degradation term $g_i$ of Eq. (\ref{Rfg}) can now be written as
\begin{equation}
g_i({\bf X},{\bf \ell}_i)=-k_i({\bf X},{\bf \ell}_i)X_i.
\end{equation}
Making the same substitution as before, $X_i=x_i+\eta_i$, and expanding to the second order on
$\eta$ leads, after ensemble averaging, to the following first order approximation:
\begin{equation}\label{Approxg}
g_i({\bf x},{\bf \ell}_i)=-\frac{K_{i}^{+}+K_{i}^{-}e^{K_i^jx_j}}{1+e^{K_i^jx_j}}x_i
\end{equation}
Combining Eqs. (\ref{fi2}) and (\ref{Approxg}), the influence function in Eq. (\ref{GeneralEq})
becomes
\begin{eqnarray}\label{Rfin}
R_i({\bf x},{\bf \Gamma}_i,{\bf \ell}_i)&=&f_i({\bf x},{\bf\Gamma}_i)- k_i({\bf x},{\bf \ell}_i)x_i\nonumber\\
&=&r_i\frac{A^j_i x_j}{(1+A^j_i x_j)(1+S^j_i x_j)}-\frac{K_{i}^{+}+K_{i}^{-}e^{K_i^jx_j}}{1+e^{K_i^jx_j}}x_i.
\end{eqnarray}

\subsection{A model of gene cluster regulation}
\label{sec:3}
Given that genes with similar expression profiles cannot be differentiated on the basis of data from microarray time series,
we are led to group genes into clusters according
to the similarity of their
profiles. This compels us to rewrite our model in terms of
gene clusters and treat the deviations in concentration levels from the
cluster average as external perturbations. Let us write $x_i=\bar{X}_q+\xi_i$, where
$\xi_i$ is the deviation of gene $i$ from the average concentration $\bar{X}_q$ of cluster $q$.
This transforms all terms of the form $A_i^jx_j$ into $A_i^p\bar{X}_p+A_i^j\xi_j$, where
$A_i^p=\sum_jA_i^j$ is the sum over all
genes in cluster $p$ denoted $N_p$, and the sum over $p$ goes over all clusters $N_c$. Inserting this into Eq. (\ref{GeneralEq}), expanding
the right hand side up to first order on $\xi$, and averaging over cluster
$q$ and over the ensemble of cells yields
\begin{eqnarray}
\dot{\bar{X}}_q&=&\varphi_q({\bf\bar{X}})+\delta_q,
\label{DifEqClust}\\
\varphi_q({\bf\bar{X}})&=&F_q({\bf\bar{X}})
-K_q({\bf\bar{X}})\bar{X}_q ,
\label{varphi}
\end{eqnarray}
where ${\bf\bar{X}}=(\bar{X}_1,...\bar{X}_{N_c})$,
\begin{equation}\label{F}
F_q({\bf\bar{X}})=\frac{1}{N_q}\sum_{i=1}^{N_q}f_i({\bf\bar{X}},{\bf \Gamma}_i),
\end{equation}
\begin{equation}\label{K}
K_q({\bf\bar{X}})=\frac{1}{N_q}\sum_{i=1}^{N_q}k_i({\bf\bar{X}},{\bf \ell}_i),
\end{equation}
and
\begin{equation}\label{Delat}
\delta_q=\frac{1}{N_q}\sum_{i=1}^{N_q}
\bigg[\sum_{j=1}^{N_q}\left(\frac{\partial f_i}{\partial\xi_j}
-X_i\frac{\partial k_i}{\partial\xi_j}\right)\langle\xi_j\rangle
-k_i\langle\xi_i\rangle-\langle\dot{\xi}_i\rangle
\bigg].
\end{equation}
In what follows, we only consider the zeroth order approximation: $\delta_q=0$.

To complete our journey from genes to clusters we must reformulate
the functions under the summation in Eqs. (\ref{F}), (\ref{K}). The
concentration levels $\bar{X}_p$ already represent the cluster average, but
the parameters $r_q$,
$A_i^p$, $S_i^p$, $K_{i}^{+}$, $K_{i}^{-}$, and $K_i^p$ need to be replaced
with some effective parameters $\rho_q$, $\alpha_q^p$, $\beta_q^p$,
$\kappa_{q}^{+}$, $\kappa_{q}^{-}$, and
$\kappa_q^p$, respectively.
Even though doing this alters the local behavior of
the multivariable functions defined in Eqs. (\ref{F}) and (\ref{K}),
their general behavior will remain the same given an appropriate set of
effective parameters. In other words, if one takes several
functions that follow a certain behavior, \emph{i.e.} starting
linearly near zero and saturating for large values,
their average will produce a function with the same behavior.
A downside to this reformulation is a loss of oversight of the
connection between the original set of parameters
$(r_q,A_q^p,S_q^p,K_{q}^{+},
K_{q}^{-},K_q^p)$ and the
effective set  ${\bf a}\equiv (\rho_q,\alpha_q^p,\beta_q^p,\kappa_{q}^{+},\kappa_{q}^{-},
\kappa_q^p)$.
The above reasoning thus yields the following model:
\begin{equation}\label{ModelFinal}
F_q({\bf\bar{X}})=\rho_q\frac{\alpha_q^p\bar{X}_p}
{(1+\alpha_q^p\bar{X}_p)(1+\beta_q^p\bar{X}_p)}
\end{equation}
\begin{equation}\label{ModelFinal2}
K_q({\bf\bar{X}})=\frac{\kappa_{q}^{+}+
\kappa_{q}^{-}e^{\kappa_q^p\bar{X}_p}}{1+e^{\kappa_q^p\bar{X}_p}}
\end{equation}

This model contains a large number of parameters compared to the number of data points, which
raises the issue of overfitting. In the following sections we address this problem and
demonstrate how a lower bound on the total number of parameters necessary to
fit the data can be determined through parameter reduction.

\section{Parameter identification}
\label{sec:4}

Having formulated a dynamical model of the cluster
network we move on to identifying the unknown parameters ${\bf a}$.
A standard practice
is to define a distance function
\begin{equation}\label{d}
d({\bf a})=
\frac{1}{\sqrt{N_tN_c}}\left[\sum_{q=1}^{N_c}\sum_{t=1}^{N_t}
\left[\frac{d}{dt}\bar{X}_q(t)-\varphi_q({\bf \bar{X}},{\bf a})\right]^2\right]^{1/2},
\end{equation}
where $N_t$ is the number of time points provided by the data, and $N_c$
represents the number of clusters. In order to estimate the derivatives 
$d{\bar X}_q(t)/dt$ one must choose an interpolating function fitting the data points
(discussed further in the next section). The function $\varphi_q({\bf \bar{X}},{\bf a})$
defines the model; it is given by eq. (\ref{varphi}).
Minimizing $d({\bf a})$ with respect
to all the parameter sets ${\bf a}$ gives the model that fits the data
the best.

The minimization problem depends heavily on
the total number of parameters to be optimized and so, reducing
the number of parameters is a valuable endeavor. We can start by
noticing that if the parameter sets ${\bf a}_q$ for each cluster
$q$ are independent of each other one can minimize the parameter
function
\begin{equation}\label{dq}
d_q({\bf a}_q)=
\frac{1}{\sqrt{N_t}}\left[\sum_{t=1}^{N_t}
\left[\frac{d}{dt}{\bar X}_q(t)-\varphi_q({\bf {X}},{\bf a}_q)\right]^2\right]^{1/2},
\end{equation}
with respect to the parameters ${\bf a}_q$ for each cluster separately.
In this manner, the problem is reduced from optimizing $3 N_c(N_c+1)$ parameters in one step to optimizing $3 (N_c+1)$ parameters $N_c$ times.

\subsection{Parameter reduction}
\label{sec:5}

We now present a useful method for reducing the number of parameters
of specific types of models, namely those of the form
$\varphi_q({\bf \bar X},\omega_q^p\bar X_p, {\bf\nu}_q)$,
where at least some of the parameters, \textit{i. e.} $\omega_q^p$s, range from $-\infty$ to $+\infty$. The remaining parameters, ${\bf\nu}_q$, are positive.

The temporal profile of each cluster consists of discrete time points.
In order to determine the functions $d{\bar X}_q/dt$ one must interpolate
the data points with a smooth continuous curve.
We chose for that purpose polynomial functions of order $w$:
\begin{equation}\label{Polyn}
\bar X_q=\sum_{n=0}^wc_{qn}t^n.
\end{equation}
Inserting Eq. (\ref{Polyn}) into
$\varphi_q({\bf \bar X},\omega_q^p\bar X_p, {\bf\nu}_q)$
gives
\begin{equation}
\varphi_q({\bf \bar X},\omega_q^p\bar X_p, {\bf\nu}_q)
=\varphi_q(c_{qn}t^n,\omega_q^pc_{pn}t^n,{\bf\nu}_q)
\end{equation}
where we again sum over repeated indices. We can now define the new
matrix element $B_{qn}=\omega_q^pc_{pn}$ which allows us to
rewrite Eq. (\ref{dq})
as
\begin{eqnarray}\label{dqNew}
&&d_q(\textbf{B}_q,{\bf\nu}_q)=\nonumber\\
&&\frac{1}{\sqrt{N_t}}\left[\sum_{t=1}^{N_t}
\left[c_{qn}nt^{n-1}-
\varphi_q(c_{qn}t^n,B_{qn}t^n,{\bf\nu}_q)\right]^2\right]^{1/2}.
\end{eqnarray}
The number of parameters to be determined for each
cluster now equals $w+1$ plus the number of elements in the set
${\bf\nu}_q$. This simple procedure
reduces the number of parameters and provides a degree of control when
interpolating the data points. For instance, one may want to
use low order polynomials at the expense of good data fit in order to reduce
the number of parameters and thus avoid the problem of overfitting.

After determining the new parameters $B_{qn}$ by minimizing
Eq. (\ref{dqNew}) we can solve for the original matrix elements
$\omega_q^p$. Notice however, that if $w$ is less than the dimension of
matrix $\omega_q^p$ we end up with many different solutions
for the parameter set $(\omega_q^{i_0},...,\omega_q^{i_w})$ depending on
what integers we assign to the indices $i_0...i_w$. For instance,
if $w=4$ then we may solve for the parameter sets
($\omega_q^1,\omega_q^3,\omega_q^7,\omega_q^8,\omega_q^9)$ or
$(\omega_q^2,\omega_q^5,\omega_q^6,\omega_q^{10},\omega_q^{12})$,
and so on. Regardless of which parameter set we chose to solve for,
the fit initially determined by minimizing $d$ will be the same.
The number of parameter sets, and thus the number of solutions for
each cluster, is $N_c!/[(w+1)!(N_c-w-1)]!$. Hence the total number of ways
the clusters can be connected to give the same value of $d$ is
$[N_c!/[(w+1)!(N_c-w-1)!]]^{N_c}$. 

The advantage of this method is twofold:
firstly, it allows one to obtain all solutions for the parameters
$\omega^p_q$ which yield the same value of $d$; and
secondly, to obtain these solutions, one only has to solve a set
of linear algebraic equations for each set 
$(\omega_q^{i_0},...,\omega_q^{i_w})$, thus significantly 
reducing the computation time. In the next section we will discuss
criteria for selecting solutions most likely adopted by nature.

\section{Modeling the glucose-lactose diauxie in \textit{E. coli}}
\label{sec:6}
We chose to model the gene expression profile of \textit{E. coli}
during glucose-lactose diauxie. The DNA microarray data was collected by
Traxler \textit{et al} \cite{Traxler}.
The diauxie experiment is designed to
observe the response of an organism to environmental stress, \emph{i.e.}
starvation. In the case at hand, the \textit{E. coli} colony was exposed
to a mixture of two sugars, glucose and lactose. The initial reaction
of the colony was to feed exclusively on glucose while
steadily growing in size. Once glucose was exhausted the growth came to a halt
for a certain amount of time after which it was resumed due to the onset
of lactose consumption. 
While the exact mechanism of this metabolic switch is not
known it has been hypothesized that the gene network of the
organism becomes rewired in response to the changing environment
(decrease in glucose) in order to survive. We want to model this
metabolic transition and study how the cluster network changes with
the varying conditions of the environment.

The model we presented in the earlier section does not include any
explicit time dependence due to environmental changes. We introduce
this feature into our model according to the following observations and
the conceptual model of glucose-lactose diauxie presented in \cite{Traxler}.
The growth arrest happens very abruptly
and therefore should not be linearly proportional to the depletion
rate of the glucose. Rather, the sudden drop in the growth rate
should be the result of the glucose level crossing a certain threshold
below which a new GRN becomes active. The primary function of the new GRN
should be to rapidly decrease cell growth while continuing to feed on
glucose. During the time of growth arrest (mixed phase)
the system makes a smooth crossover
from the glucose to lactose phase in which
the cell growth is resumed again.
The system thus ends up with the GRN that is most suited for the
consumption of lactose and cell growth.
The processes just described can be represented in symbols
as
\begin{eqnarray}\label{ModelTotal}
&&\dot{\bar X}_q=h^g(t)\varphi^g_{q}({\bf \bar X})
+h^{gl}(t,n_q)[1-h^g(t)]\varphi^{1gl}_{q}({\bf \bar X})\nonumber\\
&&+[1-h^{gl}(t,n_q)][1-h^l(t)]\varphi^{2gl}_{q}({\bf \bar X})+h^l(t)\varphi^l_q({\bf \bar X}),
\end{eqnarray}
where $g$ and $l$ stand for glucose and lactose, respectively. The glucose and lactose phases are described by the models $\varphi ^g$ and $\varphi ^l$, respectively, whereas the mixed phase is a superposition of two models defined by  $\varphi _1^{1gl}$ and $\varphi _1^{2gl}$. The functions allowing for these transitions are taken to have sigmoidal shapes: 
\begin{eqnarray}\label{SigmFuns}
&&h^g(t)=\frac{1}{1+(t/\tau_g)^{m_g}},\nonumber\\
&&h^{gl}(t,n_q)=\frac{1}{1+(t/\tau_{gl})^{n_q}},\nonumber\\
&&h^l(t)=\frac{(t/\tau_l)^{m_l}}{1+(t/\tau_l)^{m_l}}.
\end{eqnarray}
Here the exponents $m_g$ and $m_l$ are positive numbers that
determine how abruptly the system transits from the glucose to
the mixed phase and from the mixed to the lactose phase, respectively.
The constants $\tau_g$ and $\tau_l$
give the points in time of the respective transitions.
In the mixed phase, the system makes a transition
from one network to another characterized by the
exponent $n_q$ and the time constant $\tau_{gl}$ which
we consider to be half way through the mixed phase.
The functions in Eq. (\ref{SigmFuns}) can be thought
of as average fractions of cells with a particular GRN. Although
they have not been derived from experimental observation,
using sigmoidal functions is a standard practice in studying
biological transitions.

\subsection{Data analysis and gene clustering}
\label{sec:7}
Until this point we have been considering
protein concentration levels as the quantity
that is  available from experimental data. However,
the DNA microarray experiments detect the presence of
mRNA molecules - the precursors of proteins. The
pathway from mRNA to proteins occurs very quickly
in prokaryotes and so, the concentrations of these two
quantities have  an
approximately linear relationship \cite{Smolen} \cite{Bolouri}. Hence the
data on the levels of mRNA can be identified with data on protein levels.

The DNA microarray experiments as they are currently
performed do not measure the absolute mRNA concentration directly.
What they measure is the intensity of light emitted by the mRNAs
after they are illuminated by a laser. The intensity $I$ is approximately
proportional to the absolute mRNA concentration $X$. The
actual relationship between $I$ and $X$ follows
a sigmoidal curve of the form $I=aX/(1+bX)$ where $a$ and $b$ are
probe specific parameters \cite{Hekstra}.
For simplicity we assume that the linear approximation $I=aX$
is sufficient for our purposes.
The DNA microarray data are usually presented in the form
\begin{equation}
Z_i=\log_2(I_i/I_{0i})\cong \log_2(a_iX_i/I_{0i})
\end{equation}
with $I_{0i}$ being some constant
background intensity or the intensity at a given time point in a well-defined environment.
The index $i$ refers to a particular gene.
Solving this expression for $X_i$ gives
\begin{equation}\label{XvsIntensity}
X_i\cong \frac{I_{0i}}{a_i}e^{Z_i\ln2}.
\end{equation}

As we discussed earlier, a standard practice in microarray data
analysis is gene clustering, to cope with the indistinguishability of groups of gene profiles. Although in principle one
may choose to cluster the mRNA concentrations $a_iX_i/I_{0i}$, it is more relevant to cluster the $Z_i$'s. The main reason for this is that the standard deviation of mRNA concentrations measured by DNA microarray techniques, due
to noise and systematic experimental errors, has been shown to grow linearly with the expression level when this level exceeds some threshold. Taking the logarithm makes these errors additive rather than multiplicative \cite{Durbin}. The clustering is thus less sensitive to the large errors on large concentrations when applied to the logarithms of the concentrations, and thus to the $Z_i$'s.

The genes are clustered on the basis of the
similarity of their temporal expression
profiles. We use for that purpose an ordinary tree-like clustering
algorithm, which starts by considering
each gene as forming a class on its own and then groups classes two by two. In each step, the two classes are merged for which the average distance
between all pairs of gene profiles, $Z_i(t)$ and $Z_j(t)$,
taken in either of the two classes, is minimum.
The distance between the gene profiles is defined as
\begin{equation}\label{ClustDist}
D_{ij}=\left[\frac{1}{N_t}\sum_{t=1}^{N_t}[Z_i(t)-Z_j(t)]^2\right]^{1/2}.
\end{equation}
The procedure stops when the average distance $<D_{ij}>$ in the newly created class
exceeds a certain threshold. We chose this threshold to be $0.45$ which leads to 12 clusters,
each represented by the average profile $\bar{Z}_q(t)$.  
This clustering method could be modified by \textit{e. g.} adding a shift or 
introducing a scaling factor in the distance function $D_{ij}$, however, for our purposes the
simplest one suffices. One could also choose different thresholds, but we limited ourselves here to a threshold giving a sufficiently low number of clusters while keeping the profiles in each cluster reasonably similar.

The variation of each profile around the mean is defined as
$Z_i=\bar Z_q+z_i$. Making an expansion on $z_i$, Eq. (\ref{XvsIntensity})
then becomes
\begin{equation}
\frac{X_i}{I_{0i}/a_i}\cong e^{\bar Z_q\ln2}(1+z_i\ln2).
\end{equation}
Inserting this expression into our model, Eq. (\ref{DifEqClust}),
simply redefines all model parameters as $C^{old}_{ij}/(aI_{0j})=C^{new}_{ij}$
and the deviation function as $\xi_i=e^{Z_q\ln2}z_i\ln2$.

The DNA microarray data is inflicted with random noise which
makes the temporal evolution of concentration levels seem
more disjunct than it actually is. In order to
alleviate this problem we apply a simple filtering procedure
to each cluster. We define the cluster average
as a linear combination of
$\bar Z_q(t_n)$ at the $n$th time point and the two neighboring
points $n-1$ and $n+1$. In symbols:
\begin{eqnarray}\label{Filter}
\bar Z_q(t_n) &\rightarrow &\bar Z_q(t_n),  \;\;\;\;\; 
\; n=1,N_t\\
\bar Z_q(t_n)&\rightarrow & \frac{1}{2}\bar Z_q(t_n)+\frac{1}{4}\bar Z_q(t_{n-1})+\frac{1}{4}\bar Z_q(t_{n+1}),\nonumber\\
&& \;\;\;\;\;\; 
\; 1<n<N_t\nonumber
\end{eqnarray}
While other filtering methods exist, this one is the simplest and has been
successfully used before \cite{Haye}.

\subsection{Criteria for network selection}
\label{sec:8}
We used the global minimization algorithms on Mathematica
to minimize the distance functions defined in Eq. (\ref{dq})
for each cluster. Since
only a few time points of the data set belong to the glucose phase,
our modeling procedure cannot be reliably applied in this
temporal region without the risk of overfitting.
For this reason, we begin our analysis with the glucose-lactose
transition phase, which we estimate to start at the
third time point, and continue with the lactose phase
passed the time point number eight until the last (seventeenth) 
time point. Hence, the number of time points in each respective phase is:
$N_t^g=3$, $N_t^{gl}=6$, and $N_t^l=10$. These time points span a few hours.

As mentioned before, the number of possible networks
which give the same fit
is very large. However, a good fit does not guarantee
that the temporal evolution of the system will be stable
with respect to the small deviations
$\Delta_q(t)={\bar X}_q(t) -\hat{X}_q(t)$, where $\hat{X}_q(t)$ is
the modeled curve and ${\bar X}_q(t)$ the interpolated data curve. 
Since our model contains terms
such as $C_q^p\bar{X}_p$, one can see that a deviation from
the interpolated curve
$\bar X_q(t)={\hat X}_q(t)+\Delta_q(t)$ will lead to
$C_q^p{\bar X}_p+C_q^p\Delta_p$. Unless $C_q^p\Delta_p$
is small it will cause the system to deviate
more and more after each iteration of the 
differential equation solving algorithm. We therefore
argue that small parameters are likely to lead to greater stability
than large parameters. While we do not give a formal
proof here we report that running simulations with
different sets of parameters do support this
argument. Although large parameter values make
the system unstable, the opposite cannot always be
said of small parameters. Once we select the solution
set with the smallest parameter values we must weed out the
ones that fit the data points poorly.
This can be done by computing the quantity
\begin{equation}\label{Crit1}
\Omega=\left[\frac{1}{N_tN_c}\sum_{n=1}^{N_t}
|\hat{\bf X}(t_n)-{\bf \bar X}(t_n)|^2\right]^{1/2},
\end{equation}
and keeping the parameter sets which give the lowest $\Omega$.

Another restriction we impose
on the possible solutions is that the system must settle
in a fixed point after some relaxation time in the
absence of external perturbations.
We argue that the fixed point
should be of the same order of magnitude as
the average vector $\langle{\bf \bar X}\rangle=(1/N_t)\sum_n{\bf \bar X}(t_n)$.
We base this assumption on the observation that
even the most abrupt changes during the diauxie experiment
lead to the log intensity levels no larger than $|Z|\approx 2$.
It is therefore reasonable to suppose that fixed points
which differ by more than one order of magnitude from
$\langle{\bf \bar X}\rangle$ are not biologically
meaningful. We quantify this criterion by defining
the scalar quantity
\begin{equation}\label{Crit2}
\chi=|\langle{\bf \bar X}\rangle-
{\bf \hat X}(t\rightarrow t_\infty)|,
\end{equation}
where $t_\infty$ was chosen to be three times the
difference between the first and the last time point.

Random mutations in the genes and GPs can be beneficial to biological systems;
however, in many cases they degrade their performance
and even become lethal. Other random variations
such as temperature, pH factor, diet change, etc.
can also hinder the phenotype of a biological system. 
All of these changes translate into the alteration of some network
connections, \textit{i.e.} parameters $\alpha^p_q$, $\beta^p_q$, and $\kappa^p_q$. 
However, survival of biological systems partly relies on
the fact that their parameters are not rigid but can
vary within a certain range (see Gutenkunst \textit{et al} \cite{Gutenkunst}). 
A system which is robust with respect to
perturbations of the network connections is therefore
well suited for survival (for more detailed discussion of robustness,
see \cite{Kitano}).  We define a parameter
robustness function
\begin{equation}\label{Crit3}
\mu=\left[\frac{1}{N_tN_cN_s}\sum_{n=1}^{N_t}
\sum_{i=1}^{N_c}\sum_{s=1}^{N_s}\left(\frac{\partial \hat X_i(t_n)}
{\partial C_s}\right)^2\right]^{1/2},
\end{equation}
where the dummy variable $s$ runs over all model
parameters $N_s$ and $C_s$ stands for a particular
parameter. The partial derivative compares a
system with one of its variables perturbed by
a small amount to the unperturbed system.
Those networks which gave the smallest value of $\mu$ were
preferentially selected as possible candidates over the
others. We should mention that small values
of $\mu$ can have two implications. Either the system
is very sensitive to only a few parameter changes, or
it is mildly sensitive to many parameter perturbations.
The former would imply that certain parameter values
must be preserved at all cost in order for the system to
function properly, while the latter necessitates that
a large alteration in one, or several, of the parameters must
occur for a significant phenotypic change.
Irrespective of which one of these scenarios takes
place, a system with the lowest $\mu$ is said to be most robust.

\section{Results and Discussion}
\label{sec:9}
The abundance of the information obtained from DNA microarrays
scales with the number of time points. Given the scarcity
of data points in the glucose phase and the fact that the
concentration levels are nearly constant
(probably because the system has reached a steady state or fixed point), we cannot obtain
reliable information about the gene network in this temporal
region. Therefore, we start at the third time point
which marks the beginning of the growth arrest.
The last three points approach another
plateau due to depletion of the lactose. Since we do not introduce
this feature to our model we stop at the fourteenth point.

Following the procedures of parameter reduction and
parameter identification detailed in the preceding sections we
found that the mixed phase requires a polynomial of order eight
to have a good interpolation between the data points (see Eq. (\ref{Polyn})).
In the lactose phase
the interpolating polynomial turned out to be of order four.
The global optimization algorithms give less reliable results
as the number of parameters grows. For this reason, we separated the problem
into two parts. First, we considered Eq. (\ref{ModelFinal2})
to be independent of the $\kappa_q^p$'s, \emph{i.e.}
$K_q({\bf \bar X})= \kappa_q^+ + \kappa_q^-$,
which leaves
Eq. (\ref{ModelFinal}) as the source of transcription control.
We then minimized the distance function Eq. (\ref{dq}) with
respect to $\rho_q$, $\kappa_q^\pm$, $\alpha_q^p$ and $\beta_q^p$, and
recorded how well it fitted the data.
Second, we set Eq. (\ref{ModelFinal}) to a constant,
\emph{i.e.} $F_q({\bf \bar X})=\rho_q$, and optimized the distance function
$d_q$ with respect to $\rho_q$, $\kappa_q^\pm$, and the
$B_{pn}$'s.

The application of this approach to the lactose phase showed that imposing the latter assumption ($F_q({\bf \bar X})=$ constant) allowed us to fit
the data orders of magnitude better than when imposing the former assumption ($K_q({\bf \bar X})=$ constant) for
clusters $1$ through $10$.
For clusters $11$ and $12$ we had to include all parameters contained
in our model and found that the only nonzero $\beta_q^p$'s
are those with the index $p$ having values
$p=4,8,12$ and $p=11$ for cluster $11$ and $12$ respectively.

Application of the procedure just outlined to the mixed
phase yielded similar results, namely, that
keeping $F_q({\bf \bar X})$, rather than $K_q({\bf \bar X})$,
constant for all clusters
gives a much better fit of the data.
However, in the mixed phase, minimization with respect to the
parameters $B_{qn}$ of Eq. (\ref{dqNew}) yielded
$\kappa_q^p$'s that were very large. Due to
this complication we resorted
to the conventional way of parameter identification, Eq. (\ref{dq}),
and optimized $d_q$ with respect to the original
parameters $\kappa_q^p$'s. The latter gave
good results while keeping the $\kappa_q^p$'s small.

\begin{figure}
\begin{center}
\begin{picture}(0,0)(0,0)
  \put(-10,115){$a)$}
  \put(-10,-10){$b)$}
\end{picture}
\includegraphics[scale=.6]{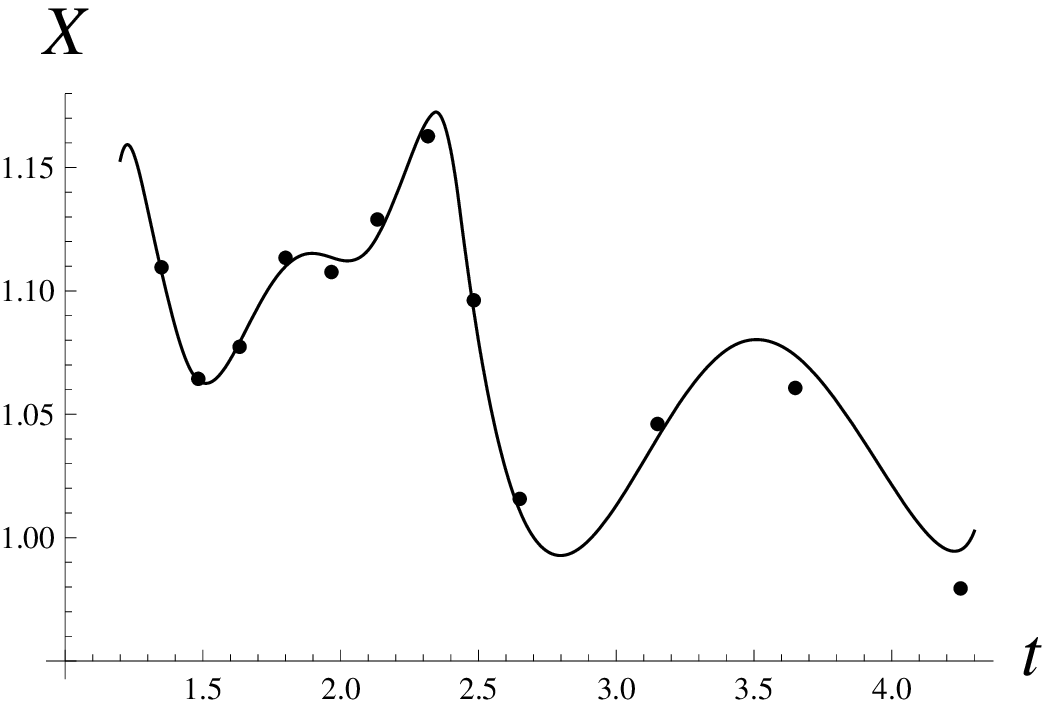}\\
\includegraphics[scale=.6]{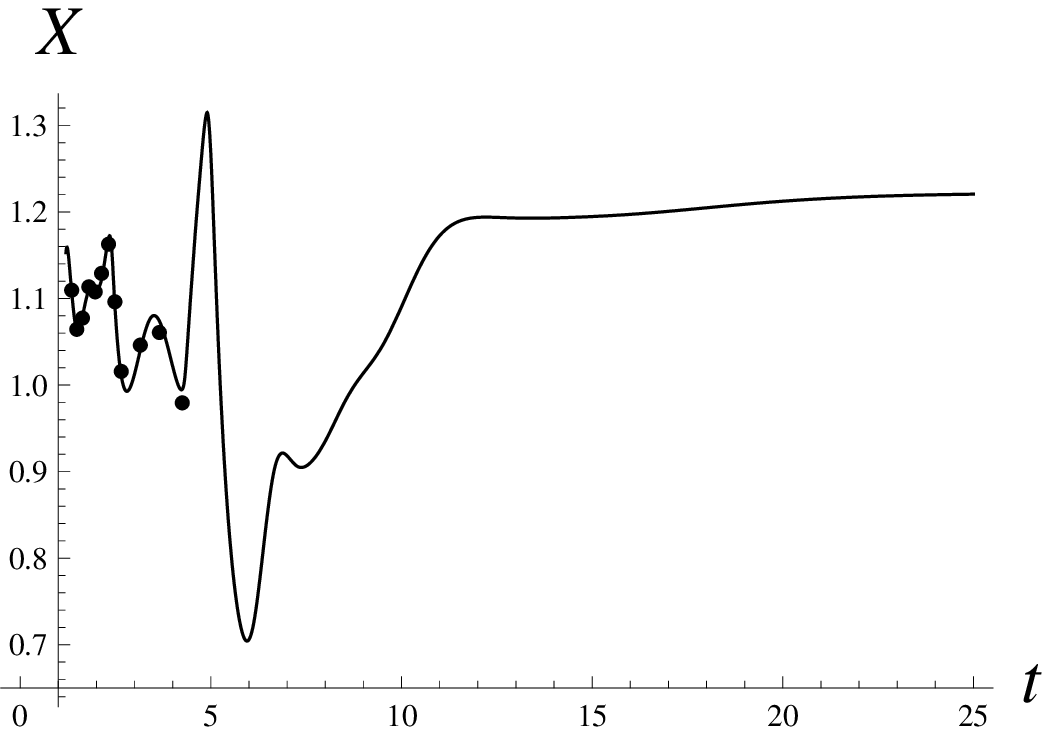}
\caption{Temporal expression profile of cluster 9.
The vertical axis is proportional to the absolute
concentration levels $\bf \bar X$, while the horizontal axis represents the time in hours. The dots and the solid line represent
the data points filtered according to Eq. (\ref{Filter}) and
the modeled curve, respectively. The transition from mixed to lactose
phase occurs at time 2.4 hours.
a) Temporal evolution during the diauxie. b) Extrapolation of the
expression profile well outside of the
experimental data showing the expression level approaching a fixed point.}
\label{Graph}
\end{center}
\end{figure}

These results suggest that the effects of
GP-GP interaction, as described by the second term of Eq. (\ref{ModelFinal}), are absolutely
necessary in gene regulation during the
glucose-lactose diauxie.
They also imply that the rate of transcription,
corresponding to the first term of Eq. (\ref{ModelFinal}),
is relatively constant indicating that
the GPs which participate in gene activation
are abundant while the ones that inhibit transcription
are low in concentration.
Another observation one can make is that
the GRN is completely
connected in the mixed phase and becomes more sparse in the
lactose phase. This means that in the mixed phase
there is no room for different parameter sets - only
one network accomplishes the temporal profiles given
by the data.

In contrast, the number of possible parameter sets in the lactose phase is very large.
In order to pick out the most probable network in the
lactose phase we
employed the network selection criteria described in the
previous section. First, we selected for each cluster the five
parameter sets  with the smallest
parameter values and then ran simulations for 350 randomly chosen
combinations
among the clusters (refer to section 3A). For each combination we computed
the three quantities $\Omega$, $\chi$ and $\mu$ (Eqs.
(\ref{Crit1}), (\ref{Crit2}), (\ref{Crit3})), which monitor the goodness of fit,
the approach to a fixed point and the robustness, respectively.
We ended up with only $10$ combinations that yielded
small values for all three criteria. Note that the fixed point criterion
$\chi$ showed a discontinuity
in the possible values it could take centering
around the numbers $\approx1$ or $\approx15$.
Figure 4 shows a three dimensional plot representing
$\Omega$, $\chi$, and $\mu$. One can see that the concentration
of points nearest to zero is relatively low. The isolated group of 10 points
within the circle comprises the best candidates for the GRN in the lactose phase.
For a particular GRN in the lactose phase, we exhibit in Figs 3 a and 3 b
the data fit of cluster 9 between the time points 1 and 14, and the
extrapolated curve showing the fixed point, respectively.
For temporal profiles of the other clusters refer to Fig 1. and Fig 2.
of the supplementary material.

Although the ``true`` GRN cannot be determined with
certainty, one can hope to at least identify the connections
that are indispensable. By comparing different possible GRNs
we can assign more importance to the connections that
appear most often. We define the average connectivity:
\begin{equation}\label{AvConn}
\langle\kappa_q^p\rangle=\frac{1}{M}\sum_n\kappa_{nq}^p
\end{equation}
where $M=10$ is the number of GRNs considered and $\kappa_{nq}^p$ is
the connection between clusters $p$ and $q$ (see Eq. (\ref{ModelFinal2})) given by the $n$th GRN.
The associated standard deviation is given by:
\begin{equation}\label{AvConnDev}
\sigma_{qp}=\left[\frac{1}{M}\sum_{n=1}^M(\langle\kappa_q^p\rangle
-\kappa_{nq}^p)^2\right]^{1/2}
\end{equation}
between clusters $q$ and $p$. If a connection
$\langle\kappa_q^p\rangle$ has a large value its
contribution to gene regulation is significant.
However, if $\sigma_{qp}$
is also large, \emph{i.e.} $\sigma_{qp}\sim
\langle\kappa_q^p\rangle$,
the certainty of this connection's value is
low and one cannot consider its
significance with confidence.
Another important
common factor is the uniformity of the sign for
each connection. If a connection has a positive
sign in one solution, it should have the same sign in
all the other solutions.

To have an objective measure of
how important a connection is, on the basis of its strength, standard deviation and sign, we define a
significance factor which ranges from 0 to 1:
\begin{equation}\label{SigFact}
S_{qp}=\frac{|\langle\kappa_q^p\rangle|}
{|\langle\kappa_{max}\rangle|}
(1-e^{-|\langle\kappa_q^p\rangle|/\sigma_{qp}})e^{-N_\pm},
\end{equation}
where $|\langle\kappa_{max}\rangle|$ is the value of the largest
average connection and $N_\pm$ is the number of times a connection
changes sign. In figure 4 we showed the gene network in the
lactose phase, with only the significant
connections indicated, as defined by $S_{qp}\geq 0.1$.

\begin{figure}
\begin{center}
\begin{picture}(0,0)(0,0)
  \put(-1.0,50){\begin{Huge}\textcircled{}\end{Huge}}
\end{picture}
\includegraphics[scale=.6]{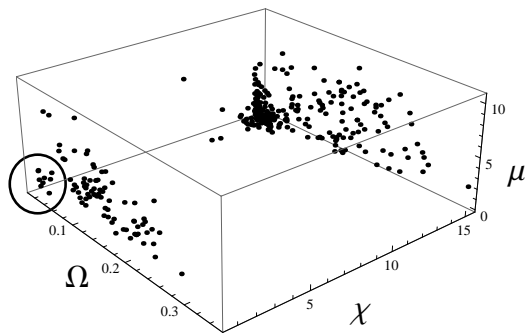}
\label{Criteria}\
\caption{Plot of $\Omega$, $\chi$, and $\mu$.
The points within the circle correspond to the best candidates
for the ''true`` GRN.}
\end{center}
\end{figure}

\subsection{Concluding remarks and outlook}
\label{sec:10}
We have presented a detailed analysis of the problem
of GRN inference through the design of a model which captures
the biochemical effects between
genes and GPs as well as the interaction
among the GPs themselves.
We hypothesized that the most important role
of the GP-GP interaction is
to vary (increase or decrease)
the characteristic time during which a GP
can perform its function. The agreement
between data and simulation based on our model
suggests that the role of the interaction among GPs
is essential in GRNs. To our surprise,
the regulation of genes
by direct binding of GPs to the genes' promoters, as described by Eq. (\ref{fi2}),
amounted to a constant independent of time in all
except two clusters, 11 and 12, in the
lactose phase. Although one may be tempted to conclude from this result that
the transcription rates of nearly all genes are constant in both phases, leaving the
non-constant degradation rate in charge of the gene regulation, it should be
kept in mind that Eqs (\ref{ModelFinal}) and (\ref{ModelFinal2}) 
deal with gene clusters, not individual genes. 
The transcription rates of all genes in a particular cluster 
may exhibit temporal variations while yielding a constant value
when averaged over the cluster.
Therefore, our results must be interpreted in the context of 
cluster network and cannot be directly compared to data on
networks containing individual genes.

\begin{figure}
\begin{center}
\begin{picture}(0,0)(0,0)
\end{picture}
\includegraphics[scale=.4]{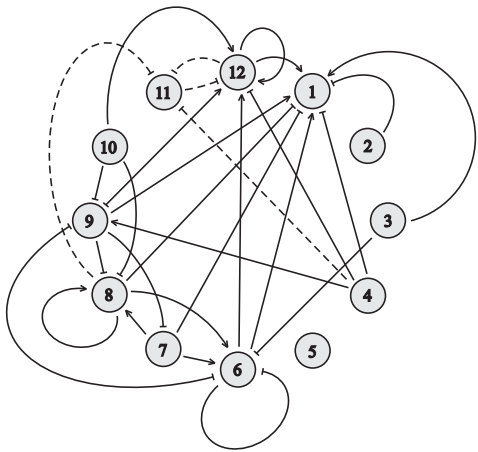}
\caption{A plausible gene cluster network in the lactose phase based
on our criteria. The full lines correspond to the $\kappa^p_q$'s of Eq. (\ref{ModelFinal2})
while the dashed lines represent the $\alpha^p_q$s 
and $\beta^p_q$s in Eq. (\ref{ModelFinal}).}
\label{grn}
\end{center}
\end{figure}

The cluster network in the lactose phase is
very sparse compared to that in the mixed phase. Previous
works on the dynamic robustness of GRNs suggests
that biological networks with low connectivity are better suited
for survival than more densely
connected networks \cite{Leclerc}.
Our results suggest that under external
stress, \emph{e.g.} starvation, the GRN of \textit{E. coli}
becomes highly connected in order to
adapt to the suboptimal conditions.
This implies that while in the mixed phase,
\emph{E. coli} is more vulnerable to random external perturbations,
upon transition to the lactose phase the robustness
with respect to environmental insults becomes restored.

The complete connectivity of the mixed phase can also be taken to mean that upon depletion of
glucose the different cells try different GRNs, each of which
is possibly sparse \cite{Tigges}. Under this assumption,
the DNA microarray data would correspond to a superimposition of different GRNs
experimented by the system until it finds the right GRN, which allows it to feed on lactose.
More experimental and theoretical work will be needed to settle this issue.

\vspace*{0.2cm}
{\bf Acknowledgements}
We thank E. Bojilova, T. Konopka, and J. M. Kwasigroch
for useful discussions. We acknowledge support from the Belgian State Science Policy Office
through an Interuniversity Attraction Poles Programme (DYSCO), and
the Belgian Fund for Scientific Research (FNRS) through a FRFC project. MR is Research Director at the FNRS.

\end{document}